\documentclass[sn-mathphys,Numbered]{sn-jnl}

\usepackage{graphicx}%
\usepackage{multirow}%
\usepackage{amsmath,amssymb,amsfonts}%
\usepackage{amsthm}%
\usepackage{mathrsfs}%
\usepackage[title]{appendix}%
\usepackage{xcolor}%
\usepackage{textcomp}%
\usepackage{manyfoot}%
\usepackage{booktabs}%
\usepackage{algorithm}%
\usepackage{algorithmicx}%
\usepackage{algpseudocode}%
\usepackage{listings}%
\usepackage{natbib}

\newtheorem{definition}{Definition}%

\title{Toward Computing Bounds for Ramsey Numbers Using Quantum Annealing}

\author[1]{Joel E. Pion}
\author[2]{Susan M. Mniszewski}
\affil[1]{University of California, Santa Barbara}
\affil[2]{Los Alamos National Laboratory}

\keywords{Monochromatic Triangle Problem, Ramsey Theory, Quantum Annealing, Optimization}

\begin{document}

\flushbottom
\maketitle
\thispagestyle{empty}

\begin{abstract}
QQuantum annealing is a powerful tool for solving and approximating combinatorial optimization problems such as graph partitioning, community detection, centrality, routing problems, and more. In this paper we explore the use of quantum annealing as a tool for use in exploring combinatorial mathematics research problems.
We consider the monochromatic triangle problem and the Ramsey number problem, both examples of graph coloring.
Conversion to quadratic unconstrained binary optimization (QUBO) form is required to run on quantum hardware. While the monochromatic triangle problem is quadratic by nature, the Ramsey number problem requires the use of order reduction methods for a quadratic formulation. The goal is to provide a method for producing special colorings of graphs which if successful would provide lower bounds for certain Ramsey numbers. We discuss implementations, limitations, and results when running on the D-Wave Advantage quantum annealer.
\end{abstract}

\section{Introduction}
\label{introduction}







Combinatorial optimization problems such as graph partitioning~\cite{Ushijima2017}, community detection~\cite{Negre2020}, centrality~\cite{Akrobotu2022}, and routing~\cite{Pion2023} are NP-complete problems that can be solved approximately using quantum-classical or quantum-only methods with a quantum annealer (QA). In this paper we explore using QAs as a tool in combinatorial mathematics research.
We provide a method for solving the monochromatic triangle (MCT) problem framed as a quadratic unconstrained binary optimization (QUBO) problem. This is followed by a description of the Ramsey number problem. MCT is quadratic, while the Ramsey problem can be higher order. A novel technique for order reduction is presented that is applicable to Ramsey problems.

We explore the effectiveness of quantum annealing on both the MCT and Ramsey problems. Both problems are important in their own right. MCT is a basic building block to the wider theory of coloring problems. Ramsey theory on the other hand is a specific coloring problem which, to the uninitiated, has a surprising number of applications ranging from number theory to network design~\cite{roberts1984applications,Gasarch_2023}. Applications of quantum computing to Ramsey theory problems have been studied for some small Ramsey numbers on an annealing device in \cite{PhysRevLett.111.130505} and have been numerically simulated for some small Ramsey numbers using adiabatic quantum evolution in \cite{Gaitan_2012}. 
To summarize the approach described in this paper, the standard method was taken for converting the coloring problem into a polynomial unconstrained optimization problem. One of the main novelties of the approach in this paper comes from the manner of reducing the polynomial to a quadratic, thus allowing quantum annealing to be applied. The order reduction method is built from a technique for solving the more basic MCT problem using quantum annealing. The aim of this method is to produce example colorings of the graph which improve the lower bound for unknown Ramsey numbers.



Quantum annealing 
uses quantum-mechanical effects such as superposition, entanglement and tunneling to find the minimum of an Ising spin glass problem. In this model each bit (or qubit) in the solution
string is represented as a {±1}. In this work however we view everything under the
QUBO perspective which is equivalent to the Ising model via a linear transformation of variables.

The QUBO formulation 
can be written as $$\min\limits_{\Vec{x}}\vec{x}^\intercal Q\vec{x},$$
where $\vec{x}$ is a vector of binary variables, $\{0,1\}$, and $Q$ is a real symmetric matrix. Upon consideration of this form one may realize the problem is to find the minimization of a multi-variable quadratic polynomial given that only binary inputs are allowed, hence the name. Quantum annealing devices are designed to minimize QUBOs 
or at least find good approximations of the minimum energy. The problem of minimizing a QUBO is NP-Hard \cite{Lewis}. This means that many challenging problems may easily be reframed as a QUBO and thus be solved using a QA. The challenge becomes then to find efficient ways to restate whatever the question of interest is as a QUBO. 

The D-Wave Advantage 4.1 QA was used in this work.
It has more than $5,000$ qubits and $35,000$ couplers arranged in a Pegasus topology. This allows for a maximum fully connected graph of 177 logical qubits or variables \cite{McGeoch}. 
Some problems were able to run directly on the quantum device. However, due to the increasing size of the problems, a quantum-classical approach was required most of the time. D-Wave’s LEAP hybrid solver~\cite{HybridHardware,HSS,HybridSolvers} with the Advantage 4.1 device was
used in this case. This is due to the limitation on the connectivity between qubits. When embedding
problems on quantum annealing hardware beyond a certain size, multiple physical qubits (as chains) may be 
required to represent logical variables. The embedding performed by the Minorminer tool~\cite{MinorMiner} is dependent on the shape of the problem and the topology of the quantum hardware. 
A time limit can be specified for the hybrid solver runtime per problem.


Order reduction in the context of this work is the concept of taking a polynomial unconstrained binary optimization problem and reducing its degree, generally until it is quadratic. This is important as the ``natural" way to convert some problems to an unconstrained binary optimization problem is inherently of higher order. And quantum annealing, as well as some classical techniques, work only with unconstrained binary optimization problems which are at most quadratic. There exists a standard procedure for order reduction, sometimes referred to as the Rosenberg technique~\cite{rosenPoly}, which works as follows.

Suppose one had the unconstrained binary optimization problem $x_1x_2x_3$. This is cubic and so to use a quantum annealing device we need to reduce it to quadratic. The idea is to replace $x_1x_2$ with an ancillary variable $a$. To do so, we need to guarantee that $a = x_1x_2$. This is achieved using the polynomial $$x_1x_2 - 2ax_1 - 2ax_2 + 3a,$$ which is $0$ whenever $a = x_1x_2$ and $>0$ whenever $a \neq x_1x_2$. One should then multiply $x_1x_2 - 2ax_1 - 2ax_2 + 3a$ by a constraint weight which is larger than the minimal achievable value of the original optimization problem. For degrees larger than $3$ one just iterates the above process. For $x_1x_2x_3...x_n$ one would use $a_1 = x_1x_2$, $a_2 = a_1x_3$ and so on in the same way as above. Additionally one should minimize the number of ancillary variables added in this way by choosing which variables to pair up carefully.

The above introduction gives a brief look into how quantum annealing works from a practical level as well as an introduction to the standard order reduction techniques when making QUBOs. In the Methods section, 
definitions and terminology for understanding coloring problems are provided. 
The MCT problem is described in more detail along with showing our novel approach on how to solve the problem using a QA. In the following section on Ramsey theory 
we introduce the standard approach to turning Ramsey problems into polynomial binary optimization problems. Then 
describe an original method for order reduction for this problem inspired by the solution to the MCT problem. After outlining the main ideas 
some ways are provided to make the method more efficient. In section \ref{theend} the results and conclusions made during the experiments are discussed. 

\section{Methods}
\label{Methods}

\subsection{Coloring Problem Definitions}
\label{definitions}

Graph coloring problems, the kind of graphs with edges and vertices, have an illustrious history and are still an important area of research for mathematicians, computer scientists, and natural scientists alike. Coloring problems have found applications in scheduling, cartography, clustering, as well as other areas \cite{colorApplication}. The two kinds of graph coloring problems relevant to this paper, which happen to both be edge coloring problems, are the monochromatic triangle problem (MCT) and Ramsey theory. Note all graphs in this paper will be assumed to be undirected.

The MCT is an NP-complete decision problem~\cite{monoHard} which is defined as follows. 

\begin{definition}
    (Monochromatic Triangle Problem): Given a graph $G$, does there exist a $2$-coloring of the edges of $G$ with no monochromatic triangle.
\end{definition}

To make sure all readers are on the same page in terms of terminology, we define a few of the above terms below.

\begin{definition}
    ($n$-coloring): Let $G$ be a graph. An $n$-coloring of $G$ is a function $$c: G_{\text{edges}} \to \{0,...,n-1\}$$ 
\end{definition}

Note that in the case of a $2$-coloring we will sometimes use `red' to refer to `$0$' and `blue' to refer to `$1$.' Additionally when not otherwise stated, assume a coloring refers to a $2$-coloring.

\begin{definition}
    (Monochromatic Triangle): Let $G$ be a graph. Let $V$ be the vertices of $G$. Let $c$ be a coloring of $G$. A monochromatic triangle is a triple of vertices in $V$, $i, j, k$, none equal, such that $(i,j),(j,k)$, and $(k,i)$ are all edges in $G$ and $c((i,j)) = c((j,k)) = c((k,i))$.
\end{definition}


Ramsey theory is the study of how large graphs need to be to guarantee certain substructures appear. First we will define what a complete graph is and then we may define classical Ramsey numbers. We shall also provide some definitions for one kind of generalization of a Ramsey number.

\begin{definition}
    (Complete Graph): A graph, $G$, is complete if for all vertices in $G$, $v,w$, with $v\neq w$, G has the edge $(v,w)$. Note the complete graph on $n$ vertices shall be denoted $K_n$.
\end{definition}

One should note that a monochromatic triangle is the same as what we refer to as a monochromatic $K_3$ and that a monochromatic $K_n$ is defined analogously.

\begin{definition}
    (Classical Ramsey Number): Let $n$ be a positive integer. Then we define and denote the corresponding Ramsey number as $$R(n) = \min\{m\in\mathbb{N} |\text{ for all }2\text{-colorings of }K_m\text{ there is a monochromatic }K_n\}.$$
\end{definition}

$R(n)$, often denoted in the literature as $R(n,n)$, is well-defined as Ramsey's theorem tells us that $R(n)$ is finite for any $n\in\mathbb{N}$ \cite{origRamsey}. Ramsey numbers are notoriously difficult to compute with a famous quote by Erd\H{o}s about how humanity would have a greater chance fighting a more advanced alien species than computing certain Ramsey numbers. Only a few Ramsey numbers are known, however, $R(1) = 1, R(2) = 2, R(3) = 6,$ and $R(4) = 18$. There are known upper bounds and lower bounds for all other Ramsey numbers, some more refined than others. For example, $R(5)$ is currently known to lie between $43$ and $48$. There are also less standard Ramsey numbers, some of which are much easier to compute than the standard ones and some of which are still quite difficult. Let us give another definition.

\begin{definition}
    (General Ramsey Number): Let $G$ be a graph. Then we define and denote the corresponding Ramsey Number as the following. $$R(G) = \min\{m\in\mathbb{N} |\text{ for all }2\text{-colorings of }K_m\text{ there is a monochromatic }G\}.$$
\end{definition}

Note that, as always in mathematics, there are further generalizations one can make, but this is sufficient for the work in this paper.

\subsection{Monochromatic Triangle Problem}
\label{MCTsection}

The monochromatic triangle problem (MCT) turns out to have a natural mapping to a QUBO structure. Consider a graph $G$ for which one wishes to answer MCT. Then the QUBO is, $$\sum\limits_{i,j,k\in G | (i,j,k)\cong \vartriangle}(e_{(i,j)}e_{(j,k)}e_{(i,k)} + (1-e_{(i,j)})(1-e_{(j,k)})(1-e_{(i,k)})).$$ To clarify the notation, the $i,j,k$ are vertices in the graph $G$. The sum is then indexing across every triplet of vertices in $G$ who when viewed as a subgraph, with all possible edges induced by $G$, is isomorphic as a graph to $K_3$. Note while this may not appear to be a QUBO as it is written as a cubic, it is in fact a QUBO and when expanded out is equal to $$\sum\limits_{i,j,k\in G | (i,j,k)\cong \vartriangle}e_{(i,j)}e_{(j,k)} + e_{(j,k)}e_{(i,k)} + e_{(i,j)}e_{(i,k)} - e_{(i,j)} - e_{(j,k)} - e_{(i,k)} + 1.$$ The QUBO is easier to interpret in the cubic form however. Essentially for every possible triangle in $G$, the $e_{(i,j)}e_{(j,k)}e_{(i,k)}$ term will punish for an all blue coloring, while the $(1-e_{(i,j)})(1-e_{(j,k)})(1-e_{(i,k)})$ term will punish for an all red coloring. The output of the summands end up being $1$ if the relevant triangle is monochromatic and $0$ otherwise. 

Minimizing the QUBO provides an answer to MCT as if an output of $0$ is achieved then the answer is yes, a coloring exists with no monochromatic triangle. If the output is $>1$ then the answer is no, every coloring has a monochromatic triangle. Moreover, even if the answer to the MCT is no, minimizing the QUBO will tell you what is the minimum number of monochromatic triangles a coloring of the graph can produce. On top of this, minimizing the QUBO also provides the actual coloring which minimizes the number of monochromatic triangles.


Two observations about this QUBO are (1) that it does not use any particular structure from triangles beyond the number of variables and (2) that it is a naturally ``cubic" condition which is really quadratic. We will exploit (1) in section \ref{ORRT}. For (2) we should note that the ``cubic" summand used above is not unique in this feature. A sufficient and necessary condition for this to occur is that the sum of the products of the coefficients of the dominant terms is $0$. 

Lastly, the reader should note that when the authors refer to the MCT summand in later sections they refer to the following equation, $$(e_{(i,j)}e_{(j,k)}e_{(i,k)} + (1-e_{(i,j)})(1-e_{(j,k)})(1-e_{(i,k)})).$$

\section{Ramsey Theory}
\label{ORRT}


First it should be noted that QUBO solvers, quantum or classical, cannot guarantee optimal solutions in reasonable timeframes for sufficiently large problems. Approaching Ramsey problems of interest currently, through the lens of constructing a coloring, are sufficiently large, else they would not be of interest. Because of this, it must be said that any attempt of this style would have little chance of finding an exact Ramsey number. 
The primary use of this framework
is to improve lower bounds by constructing colorings which avoid certain monochromatic subgraphs. In addition one can gain empirical evidence that a Ramsey number has been found when every coloring attempted produces certain monochromatic subgraphs. The most straightforward way to write a binary optimization which searches for a coloring of $K_m$ which contains no monochromatic $K_n$, i.e. trying to show $R(n)>m$, is $$\sum\limits_{(i_1,...,i_n)\cong K_n}(\prod\limits_{r,s=1 | r<s}^ne_{(i_r,i_s)} + \prod\limits_{r,s=1 | r<s}^n(1-e_{(i_r,i_s)}).$$ This is completely analogous to what 
we did for the monochromatic triangle. The higher order polynomial formulation described here is taking the sum across every $K_n$ in $K_m$ of the product of the edges in the given $K_n$ plus the product of '$1-$ edges' in that same $K_n$. Thus in 
net the equation outputs the number of monochromatic, i.e. all $0$'s or all $1$'s, $K_n$ subgraphs in $K_m$. Importantly the equation can only output its minimum value $0$ if there are no monochromatic $K_n$ subgraphs in $K_m$. A novel part of our approach will be how we do the order reduction. Rather than use the standard method which is listed in the Introduction (section \ref{introduction}), we will use observation (1) from subsection \ref{MCTsection}. The method will be explained using $R(4)$, which ends up with degree $6$ terms. $R(n)$ for $n>4$ can be done similarly with only slight modifications.  

\subsection{Order Reduction}
\label{orderreductionssection}


\begin{algorithm}
\caption{Order Reduction for $R(4)$: Note the algorithm provided here is meant as a form of psuedo-code for implementing the order reduction described in subsection \ref{orderreductionssection}. Note in this algorithm, $edge(a,b)$ refers to a variable representing the edge between vertex $a$ and vertex $b$. While $edges((a,b),(b,c)))$ refers to an ancillary variable added which is associated to both edges listed within. Finally an index $i$ is added in edges$_i$($(a,c),(a,d)$)) to reference that some ancillary variables are associated to the same pair of vertices but are distinct from each other.}
\label{R4reductalg}
\begin{algorithmic}[1]

\Procedure{Reduced QUBO Generation}{$n$}        \Comment{$n < 18$}
    \State \texttt{QUBO $=$ $0$}
    \State $i = 0$
    \For{\texttt{$0\leq a<b<c<n-1$}} \Comment{\texttt{MCT\_QUBO() is the summand of the MCT QUBO}}
        \State \texttt{QUBO = QUBO $+$ MCT\_QUBO(edge($a$,$b$),edge($b$,$c$),edges($(a,b),(b,c)$))}
        \For{\texttt{$c<d<n$}}
            \State $i = i + 1$
            \State \texttt{QUBO = QUBO $+$ MCT\_QUBO(edge($a$,$c$),edge($a$,$d$),edges$_i$($(a,c),(a,d)$))}
            \State \texttt{QUBO = QUBO $+$ MCT\_QUBO(edge($b$,$d$),edge($c$,$d$),edges$_i$($(b,d),(c,d)$))}
        \EndFor
    \EndFor
    \State \texttt{Return QUBO}
\EndProcedure
\end{algorithmic}
\end{algorithm}

\begin{figure}[ht]
\centering
\includegraphics[width=0.7\linewidth]{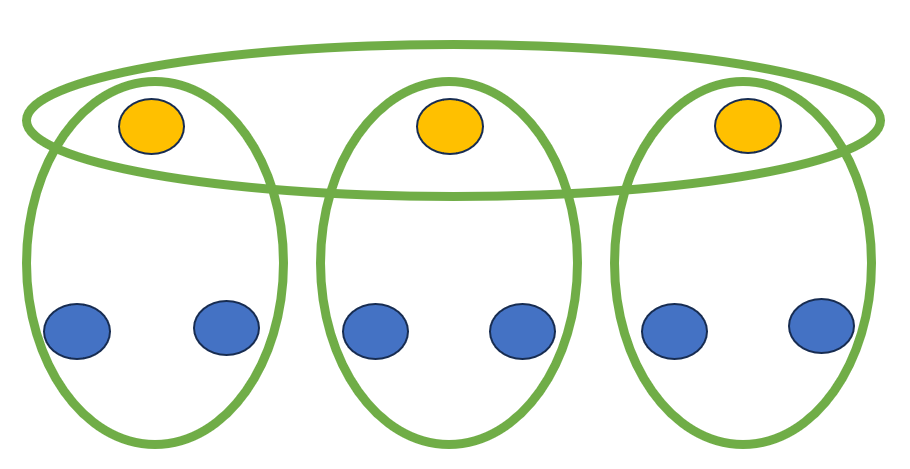}
\caption{$R(4)$ Order Reduction: Blue dots are variables directly representing edges in the graph. Yellow dots are ancilla variables added to each $K_4$. Green circles are MCT ``not all the same'' equations. The picture represents the general idea presented in this paper for order reduction in $R(4)$.}
\label{R(4) reduction}
\end{figure}

The goal of this section is to produce a method for reducing the problem of search for $K_4$-free colorings of $K_n$ from degree $4$ to degree $2$ so as to be made into a QUBO. We will use Fig.~\ref{R(4) reduction} as a reference picture. Consider the blue circles as edges, each representing one of the $6$ edges in a $K_4$. The yellow circles are ancillary variables which are added to aid in the order reduction. The green circles represent a QUBO term like the summand for the MCT QUBO. The MCT QUBO summand punishes if and only if all $3$ variables are the same. To see why the QUBO outlined in Fig.~\ref{R(4) reduction} works, let us consider the different cases. First let us consider the monochromatic case. Assume without loss of generality that all $6$ edges, the blue circles, are colored blue like they are depicted. Now let us first consider the leftmost ``do not make all $3$ the same" green circle. Since both the edges in the circle are blue, the ancillary bit must be red or else the left green circle's QUBO will output a $1$ rather than a $0$. The same logic holds similarly for the middle and right green circle. Thus all three ancillary bits are red. Thus the top green circle QUBO will output a $1$. Hence a monochromatic $K_4$ corresponds to at least a $1$. Now let us consider the other case. Without loss of generality suppose the leftmost blue circle is not equal to all the other edges in color and let us assume it is red. We know by assumptions one of the other edges is blue. First assume that a blue edge lies in the middle or right circle. In this case the leftmost ancillary variable can be blue while the middle or right ancillary variable can be red which allows the QUBO to still output $0$. If it is not the case that a blue edge lies in the middle or right green circle, then they must all be red. Hence the middle and right ancillary variables are blue. Thus the blue edge must be in the left green circle. And so the left ancillary variable may be red without punishment. Thus the QUBO may still achieve $0$.  Hence the QUBO may output a $0$ if and only if the edges are not monochromatic. 

We can make this reduction more efficient than stated. The primary downside of this order reduction method is 
the need for ancillary variables. And since there are currently three variables added for every $K_4$ in the graph, this leads to quite a large QUBO. Any reduction in variables can be helpful. The first idea will be to use how the $K_4$'s overlap as they can only overlap in specific ways. A $K_4$ overlap can involve $1, 2,$ or $3$ vertices which corresponds respectively to $0, 1$, or $3$ edge overlap. When there is a three edge overlap we can remove the need for one ancillary variable using the following method. 

\begin{figure}[ht]
\centering
\includegraphics[width=0.7\linewidth]{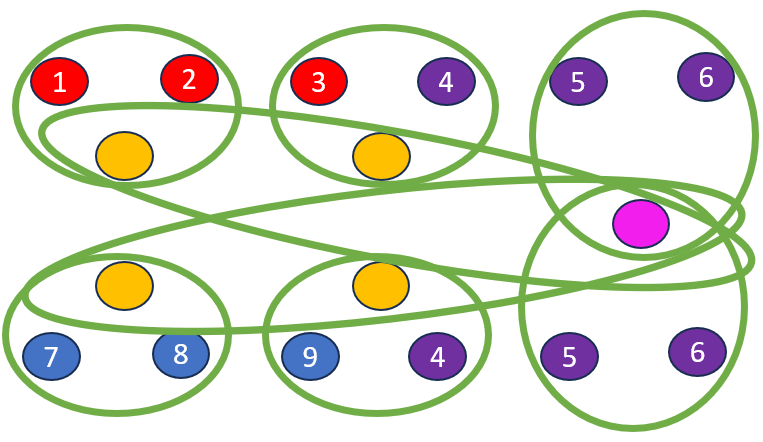}
\includegraphics[width=0.6\linewidth]{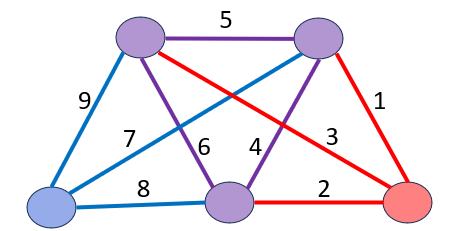}
\caption{$R(4)$ Variable Elimination. The top figure shown is an illustration for how to reduce the number of variables for the order reduction method described in this paper. The figure represents two different $K_4$'s with the red, blue, and purple dots representing variables directly associated to edges. The red, blue, and purple dots are labeled with numbers which correspond with numbers on the lower figure, showing which edges correspond to which dots in the upper figure for this example. The lower figure is an example  pair of $K_4$'s with a three vertex overlap from our graph $K_n$. The purple dots in the top figure are for edges shared between the two $K_4$'s. The yellow and pink dots represent ancillary variables added to the $K_4$. The pink dot is an ancillary variable which can be reused between certain $K_4's$. Green circles are MCT ``not all the same'' equations.}
\label{R(4) elimination}
\end{figure}

With Fig.~\ref{R(4) elimination}'s top image as reference consider the top row of red and purple dots to represent variables corresponding to the edges in the first $K_4$. The blue and purple dots will represent variables corresponding to the edges in the second $K_4$ with the purple dots being shared edges. Note the purple dots are the three dots on the right for both the top and bottom row and purple dots in the same column are the same variable. The four yellow variables in the middle middle and middle left are ancillary variables. The pink variable on the middle right is also an ancillary variable. Each green circle  encompasses three variables and represents making a ``do not make all $3$ the same" QUBO with the encompassed variables. The difference between what we had before and what we have now is that one of the ancillary variables is now shared. The only situation in which our argument for validity from before must be updated is when the two shared variables in the same 
circle have different colors. In this case the ideal is that our QUBO can still output a 0 since neither $K_4$ is monochromatic and we will show this is the case. As there is a third purple shared variable in the middle green circles, the two middle yellow ancillary variables may be the same color, whatever is opposite of the middle circle's purple variable. As the right pink ancillary variable may be either red or blue under these hypotheses, the pink ancillary variable may just be whatever is the opposite of the middle yellow ancillary variables. All the other conditions are easily satisfied. 

Applying this method by considering every $K_4$ in $K_n$ in numerical order and using this variable elimination technique to share ancillary variables whenever the first three vertices match allows us to use only $$\binom{n-1}{3} + 2\binom{n}{4} + \binom{n}{2}$$ variables. To break down this equation note the $\binom{n}{2}$ is for the variables directly representing an edge in the graph, the rest are ancilla. The $2\binom{n}{4}$ term comes out as every $K_4$ requires two new variables as described above. Finally, for each `initial $K_3$' among our $K_4$'s we will add one more variable. These variables are shared between all $K_4$'s with that `initial $K_3$'. This gives us the $\binom{n-1}{3}$ term. Note by numerical order the authors are assuming every vertex is assigned a unique number $1$ to $n$ and considering each $K_4$ as a sorted ordered tuple of vertices. This ordering is used to find when two $K_4$'s have a common initial `$K_3$.' Using the standard order reduction technique mentioned in the Introduction (section \ref{introduction}) in a similarly efficient manner by first reducing the edges in the shared triangle leads to the larger $$2\binom{n-1}{3} + 2\binom{n}{4} + \binom{n}{2}$$ variables. Once again the $\binom{n}{2}$ are variables directly representing an edge with the others being ancillary. The $2\binom{n}{4}$ is for the ancillary variable representing $x_1x_2x_3x_4$ and $x_1x_2x_3x_4x_5$ in each $K_4$. While the term $2\binom{n-1}{3}$ is for the term representing $x_1x_2$ and $x_1x_2x_3$ which are repeatable between $K_4$'s with the same $x_1,x_2,x_3$.

It is important to note that the authors' variable elimination technique is not fully optimized in its current form. A simple improvement that could be made, which is not implemented in this paper, would be to also share the `last/rightmost' ancillary variable among the $K_4$ order reduction diagrams for $K_4$'s with vertices $ij(n-1)n$ in $K_n$ for $1\leq i<j\leq n-3$. This works as in our current construction, these diagrams do not share any ancilla variables with any other of the $K_4$ diagrams. Note that our argument used to prove the validity for the construction in this paper showed that a sufficient condition for our shared ancilla construction to work is that no $K_4$ diagram has more than one shared ancilla. Notably though, this is not a necessary condition. This is notable since for $K_5$, where the main construction from this paper would use $14$ ancilla, the authors found a construction which used only $11$ ancilla; however, certain $K_4$ diagrams had more than one shared ancilla. On the other hand, there are also constructions with $K_4$ diagrams that have multiple shared ancilla which are not valid. The authors are not yet able to derive an elegant or simple way to determine which constructions will be valid and as a consequence do not have a way to generalize such constructions to larger $K_n$. A good future research question, and interesting combinatorial question, would be to find, using this framework, the minimal number of necessary ancilla for a given $K_n$.
\subsection{Extending the method to $R(5)$ and Beyond}
\label{extending}

\begin{figure}[ht]
\centering
\includegraphics[width=0.7\linewidth]{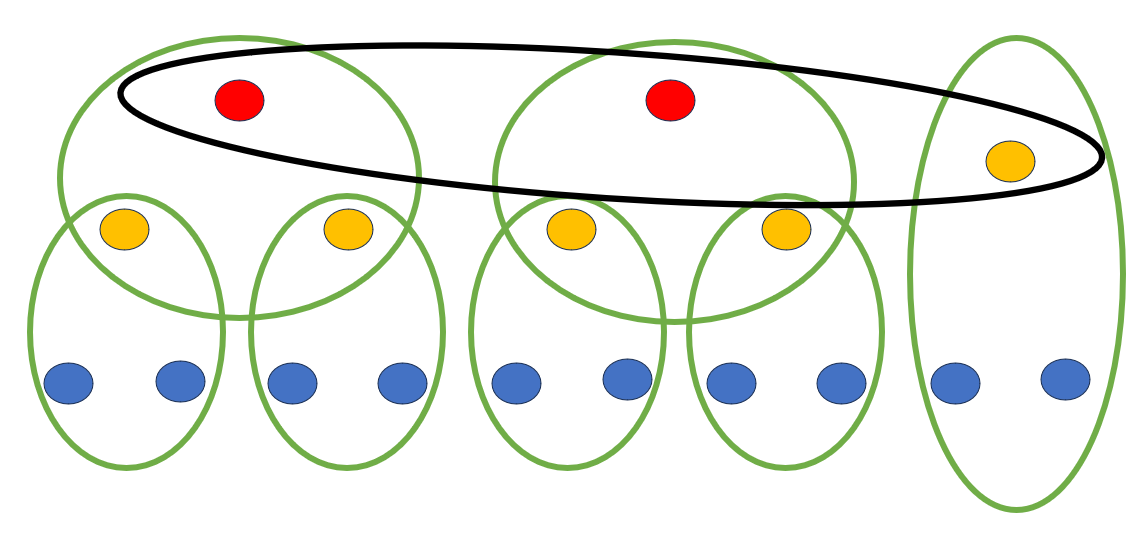}
\caption{$R(5)$ Order Reduction: Blue dots are variables directly representing edges in the graph. Yellow dots are ancilla variables added to each $K_5$. Red dots are further ancilla variables added. Green circles are MCT ``not all the same'' equations. Black circles are slightly modified MCT ``not all the same'' where the yellow dot variable is inverted, i.e sent to one minus itself. This means the black circle only punishes if the two red dots have the same output but a different output than the yellow dot in the circle.  The picture represents the general idea presented in this paper for order reduction in $R(5)$.}
\label{R(5) reduction}
\end{figure}

\begin{figure}[ht]
\centering
\includegraphics[width=0.7\linewidth]{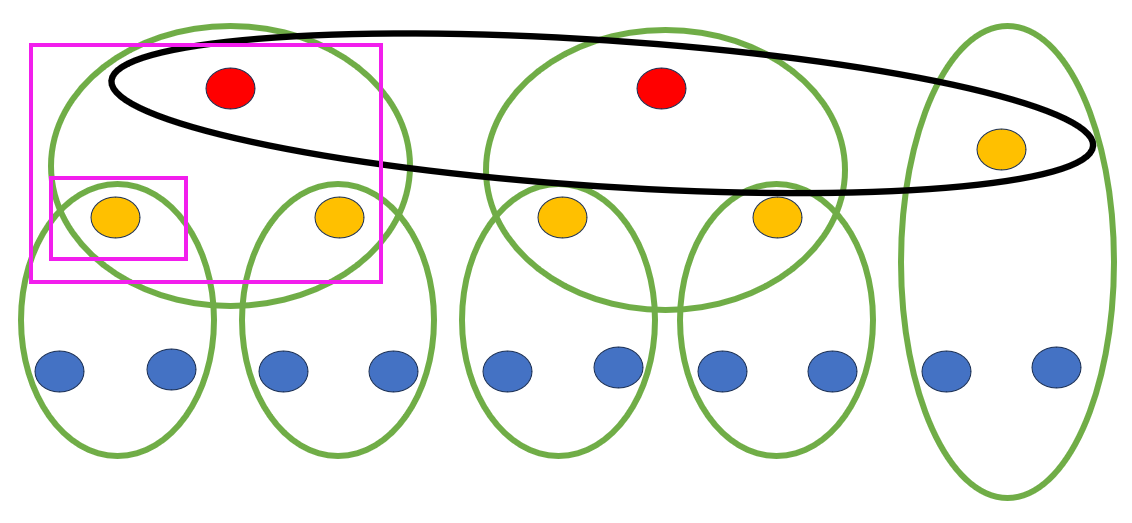}
\caption{$R(5)$ Blue dots are variables directly representing edges in the graph. Yellow dots are ancilla variables added to each $K_5$. Red dots are further ancilla variables added. Green circles are MCT ``not all the same'' equations. Black circles are slightly modified MCT ``not all the same'' where the yellow dot variable is inverted, i.e sent to one minus itself. This means the black circle only punishes if the two red dots have the same output but a different output than the yellow dot in the circle. The picture represents the general idea presented in reducing the number of variables used in $R(5)$.}
\label{R(5) elimination}
\end{figure}

The key to extending the method described for $R(4)$ to $R(n)$, $n\geq4$ is best understood by considering visually how the $R(4)$ case in Fig.~\ref{R(4) reduction} and Fig.~\ref{R(4) elimination} extends to the $R(5)$ case in Fig.~\ref{R(5) reduction} and Fig.~\ref{R(5) elimination}. Let us start with Fig.~\ref{R(5) reduction}. The idea is to pair the edges as best we can, sometimes we will end up with an odd number of edges, and add one ancillary variable for each pair. Then with any leftover edges and the newly created ancillary variables we again pair them as best we can and add one more ancillary variable for each pair. This process is 
repeated until we end with precisely 3 variables. We can think of the variables as lying on levels, with edge variables being level $0$ and each ancilla variable added is, in a recursive manner, one level higher than the highest level of the pair inducing it. Then for this pair and the newly created ancilla variable, we apply an MCT relation on the trio, potentially a modified MCT relation replacing the lower level of the inducing pair in the trio with its complement, i.e. ``$1$ minus the variables'', if it differs from the higher level of the pair by an odd number. Then for the final $3$ variables we also apply an MCT relation, and, using the highest level variable in the trio for reference, we may modify the other two variables in this MCT relation to their complements if they differ from the top level variable in the trio by an odd number of levels.

Then Fig. \ref{R(5) elimination} provides a visual reference for how to reduce the number of variables used in the generalized case when comparing to the $R(4)$ case. The pink boxes in Fig.~\ref{R(5) elimination} correspond to shareable ancilla variables between various $K_i$. The way this would work for a general $R(n)$ is as follows. Assume we associate to each vertex of $K_m$ a number, where $m$ is the number of vertices in the graph. Then we list the edges filling in sub-$K_i$'s of $K_n$ for $3\leq i\leq n$ in lexicographic order when considered as pairs of vertices from the $K_i$ until all edges in the $K_i$ are used. From here, similar to what we did with $R(4)$, one can reuse all ancillary variables for $K_n$'s supported completely by variables representing edges, except the last such edge, from $K_i$ for other $K_n$'s using the same first $i$ edges. The reason the last such variable corresponding to an edge must be ignored is to make sure there is an ancillary variable attached to a shared edge that provides one degree of freedom to avoid contradicting coloring requirement on the ancillaries.

\subsection{Pre-coloring}
\label{precoloring}

\begin{figure}[ht]
\centering
\includegraphics[width=0.7\linewidth]{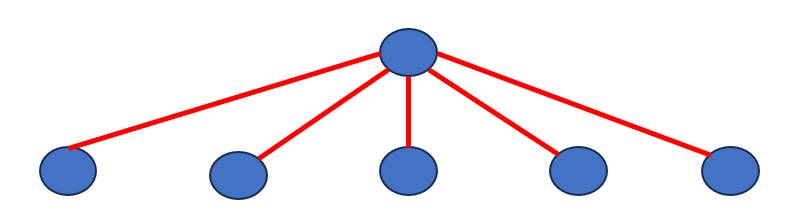}
\caption{Star Graph: $S_5$. Pictured here to help visualize pre-coloring.}
\label{star graph}
\end{figure}

Another technique for variable reduction is to pre-color some part of the graph. The gist of the idea is to exploit known Ramsey numbers and the symmetry of $K_n$. There are several ways to do this. One way is to use a smaller $K_i$, another is with star graphs. Star graphs have a central vertex and then some number of vertices radiating out from the central one. This can be seen in Fig.~\ref{star graph} where the central vertex is the top one. 

The standard notation for a star graph is $S_n$ where $n$ is the number of vertices radiating outward, so $S_n$ has $n+1$ vertices. Using the well-known result $$R(S_n) =   \left\{\begin{array}{ll} 2n-1 & 2|n \\ 2n & \text{else} \\ \end{array} \right.$$ and taking for example $K_{17}$, the target graph to color for $R(4)$, we find that any coloring must have a monochromatic $S_8$. Recall from subsection \ref{definitions} that $R(S_n)$ means we want that smallest complete graph which has a $2$-coloring without a monochromatic $S_n$. Thus our goal coloring of $K_{17}$ must have a monochromatic $S_8.$ As $K_n$ is completely symmetric we may choose where $S_8$ goes. If we pick $S_8$'s center to be vertex $16$ and then connect vertex $16$ to vertices $0$-$7$ we will find good synergy with our previous variable elimination technique, eliminating in this case a further $$8 + 8\binom{8}{2} + \binom{8}{3}$$ 
variables. To understand this number, observe the first $8$ in the equation corresponds to the variables representing edges we pre-color. The $8\binom{8}{2}$ is interpreted as the number of $K_4$'s in $K_{17}$ such that two vertices in the $K_4$ come from vertices $0-7$, one comes from vertices $8-15$, and also includes vertex $16$. For each such instance, exactly one ancillary qubit can be pre-colored. Lastly the $\binom{8}{3}$ is similar in that it corresponds to the number of $K_4$'s in $K_{17}$ with three edges from the vertices $0-7$ and includes vertex $16$. Once again in each such instance an ancillary qubit can be pre-colored. We can choose the $S_8$ to be red which can be done as the choice between colors is currently completely symmetric. This technique may then be repeated with only a slight alteration. Consider the remaining $9$ vertices not included in the $S_8$ chosen before. We know from Ramsey theory that in the remaining $K_9$ there is a monochromatic triangle. Thus we may pick a monochromatic triangle and by symmetry may choose wherever we wish to place it in the $K_9$ by symmetry. In this situation, however, we can not just set the color since the whole graph is no longer symmetric. But since we know the triangle is monochromatic, we may replace all three variables with a single variable $x$. Then any ancillary variables connecting two of the three variables may be set to $1-x$. This type of process may continue until there are not enough vertices in the graph to guarantee any helpful monochromatic sub-graphs. Note 
the pre-coloring method as described was not fully implemented, only a single $S_8$ was pre-colored. The $S_8$ though was chosen to minimize the number of ancillary variables needed. 

Using the methods described we were able to generate $K_4$ monochromatic-free colorings of $K_{11}$ consistently, $K_{12}$ often, and occasionally $K_{13}$ using three minutes of D-Wave's LEAP hybrid solver. Note that $K_{11}$ generally needs no more than $30$ seconds. 

\section{Results and Discussion}
\label{theend}







In one experiment data was collected on how well the Advantage hardware worked with the MCT algorithm developed in this paper. Table~\ref{completeTable} is the data from this experiment. Note each table entry was tested three times. A single number is provided if each trial resulted in the same value, while all three values are provided in increasing size in all other cases. The $n$ column is for what $K_n$ was being colored. As one can see simulated annealing achieved the minimum number of triangles for every $n$ tested. The simulated annealing here and throughout this section was done using a python package~\cite{simAnnealGIT} 
with the ideal Ramsey unconstrained binary optimization equation given at the beginning of the Ramsey Theory (section \ref{ORRT}) and Order Reduction (subsection \ref{orderreductionssection}). The pure quantum method achieved optimal results up to $K_8$, on $K_{10}$, as well as one run on $K_{12}$. The quantum annealing results were never worse than $10\%$ above the minimum and beyond a certain size was generally only $5\%-6\%$ worse than the minimum. The experiment ended at $n=20$ as the embedding time at $n=21$ began to take too long. The quantum annealing experiments were run by taking $1000$ reads and taking the best sample. A chain strength of $600$ and the annealing time was set to 20 microseconds. A different embedding was used during each run.

\begin{table}
    \centering
    \begin{tabular}{cccc}
        $n$ & Minimum number of triangles &  Simulated Annealing & Pure Quantum \\
        5 & 0 & 0 & 0\\
        6 & 2 & 2 & 2\\
        7 & 4 & 4 & 4\\
        8 & 8 & 8 & 8\\
        9 & 12 & 12 & 13\\
        10 & 20 & 20 & 20\\
        11 & 28 & 28 & 29\\
        12 & 40 & 40 & 40/41/41\\
        13 & 52 & 52 & 55/55/56\\
        14 & 70 & 70 & 71/72/74\\
        15 & 88 & 88 & 93/93/94\\
        16 & 112 & 112 & 116/116/118\\
        17 & 136 & 136 & 144/144/145\\
        18 & 168 & 168 & 175/175/176\\
        19 & 200 & 200 & 209/211/212\\
        20 & 240 & 240 & 251/252/252\\
    \end{tabular}
    \caption{$K_n$ MCT quantum and simulated annealing comparison. Numbers in table represent the number of monochromatic triangles that were produced using the various methods to color $K_n$.}
    \label{completeTable}
\end{table}

When testing the MCT Problem on D-Wave's LEAP hybrid solver we were able to match the OEIS data for minimum number of monochromatic triangles for a coloring of a complete graph on $n$ vertices~\cite{monoOEIS} for graphs with at least $60$ vertices. The code was also tested and successful with several non-complete graphs as well. 
The QUBO used to solve the MCT here uses the same idea as Bian et al.~\cite{DWAVERamsey} in which a lower bound for $R(3)$ was computed.

In the next experiment we collected the data in Table \ref{halfTable} to compare MCT on non-complete graphs for simulated annealing with hybrid methods and purely quantum methods. Once again each experiment was run three times, with separate listings only if different results appeared. The purely quantum method used the best result of $2000$ samples, a chain strength of $600$, a $60$ microsecond annealing time, and $5$ reverse spin-transformations. In addition a different embedding was used on each run. These values were chosen after testing a few combinations of parameters. The graphs were randomly generated with roughly $50\%$ edge saturation on each trial. The $n$ column refers to the number of vertices in the graph. Simulated annealing was used as a baseline for the minimum number of triangles and the other methods are compared to simulated annealing and printed as (other method - simulated annealing). One can see that the simulated annealing and hybrid methods always agreed in our trials and presumably found the true minimum. Note the hybrid method was limited to $5$ seconds a run. The purely quantum method found the same coloring as the simulated annealing and hybrid method only for $n=10$ and the raw difference between answers got progressively worse as the graph grew larger. Once again the experiment was capped at $n=30$ as for larger graphs the embedding time began to grow large. 

\begin{table}
    \centering
    \begin{tabular}{cccc}
        $n$ & Hybrid & Pure Quantum ~ ~ ~ ~ ~ ~ ~ ~ ~ ~ ~ ~ ~ ~ ~ ~ ~ ~ ~ ~ ~ ~ ~ ~ ~ ~ ~ ~ ~\\
         10 & 0 & 0~ ~ ~ ~ ~ ~ ~ ~ ~ ~ ~ ~ ~ ~ ~ ~ ~ ~ ~ ~ ~ ~ ~ ~ ~ ~ ~ ~ ~\\
         15 & 0 & 3/4/13~ ~ ~ ~ ~ ~ ~ ~ ~ ~ ~ ~ ~ ~ ~ ~ ~ ~ ~ ~ ~ ~ ~ ~ ~ ~ ~ ~ ~\\
         20 & 0 & 16/17/19~ ~ ~ ~ ~ ~ ~ ~ ~ ~ ~ ~ ~ ~ ~ ~ ~ ~ ~ ~ ~ ~ ~ ~ ~ ~ ~ ~ ~\\
         25 & 0 & 30/40/48~ ~ ~ ~ ~ ~ ~ ~ ~ ~ ~ ~ ~ ~ ~ ~ ~ ~ ~ ~ ~ ~ ~ ~ ~ ~ ~ ~ ~\\
         30 & 0 & 78/85/101~ ~ ~ ~ ~ ~ ~ ~ ~ ~ ~ ~ ~ ~ ~ ~ ~ ~ ~ ~ ~ ~ ~ ~ ~ ~ ~ ~ ~\\
    \end{tabular}
    \caption{50\% edge saturation MCT comparison. Numbers in the table represent the difference between the number of monochromatic triangles produced by the column's method versus simulated annealing on a random graph with $n$ edges and roughly $50\%$ edge saturation.}
    \label{halfTable}
\end{table}

The purely quantum methods were studied to understand how they performed when running the $R(4)$ Ramsey problem. The experiments were run by taking $1000$ reads and taking the best sample. A chain strength of $600$ and an annealing time of 20 microseconds were used. A different embedding was used for each run. Similar to before these parameters were chosen after a brief parameter search. Table \ref{K4quantum} is the result. The $n$ column refers to the $n$ in $K_n$. And again, each trial was run three times. As $R(4) = 18$ the minimum value for each row is $0$, however, the purely quantum runs only found this for $K_6$. The number of monochromatic triangles found increased with $n$ as is to be expected as the size of the problem grows. 

\begin{table}
    \centering
    \begin{tabular}{cc}
        $n$ & Monochromatic $K_4$'s ~ ~ ~ ~ ~ ~ ~ ~ ~ ~ ~ ~ ~ ~ ~ ~ ~ ~ ~ ~ ~ ~ ~  ~ ~ ~ ~ ~ ~ ~ ~ ~ ~ ~ ~ ~ ~ ~\\
        6 & 0~ ~ ~ ~ ~ ~ ~ ~ ~ ~ ~ ~ ~ ~ ~ ~ ~ ~ ~ ~ ~ ~ ~  ~ ~ ~ ~ ~ ~ ~ ~ ~ ~ ~ ~ ~ ~ ~\\
        7 & 1/1/3~ ~ ~ ~ ~ ~ ~ ~ ~ ~ ~ ~ ~ ~ ~ ~ ~ ~ ~ ~ ~ ~ ~  ~ ~ ~ ~ ~ ~ ~ ~ ~ ~ ~ ~ ~ ~ ~\\
        8 & 2/2/3~ ~ ~ ~ ~ ~ ~ ~ ~ ~ ~ ~ ~ ~ ~ ~ ~ ~ ~ ~ ~ ~ ~  ~ ~ ~ ~ ~ ~ ~ ~ ~ ~ ~ ~ ~ ~ ~\\
        9 & 3/7/8~ ~ ~ ~ ~ ~ ~ ~ ~ ~ ~ ~ ~ ~ ~ ~ ~ ~ ~ ~ ~ ~ ~  ~ ~ ~ ~ ~ ~ ~ ~ ~ ~ ~ ~ ~ ~ ~\\
    \end{tabular}
    \caption{Pure Quantum $R(4)$ computations. Numbers represent the number of monochromatic $K_4$'s created from coloring $K_n$.}
    \label{K4quantum}
\end{table}

Initially, the goal of this paper was to attempt to develop a method using quantum annealing to improve the lower bounds for classical Ramsey numbers. Using current quantum and quantum-classical hardware and software with the method described in this paper to compute the greatest lower bound for $R(4)$ was unsuccessful as there were not enough qubits and connectivity. This leaves the loftier goal of improving the lower bound for $R(5)$ or $R(6)$ still out of reach. However, the methods were able to produce $K_4$-free colorings of $K_{13}$ as well as finding the maximal $K_3$-free coloring for $R(3)$. Our methods were unable to proceed past $K_{14}$ for $K_4$-free colorings. The simulated annealing had little issue finding $K_4$ monochromatic-free colorings of $K_{15}$ if given a minute or two, although it was unable to do so for $K_{16}$ or $K_{17}$ even with five or ten minutes. The fact that the simulated annealing outperformed D-Wave's LEAP hybrid solver
is indicative of the failure of the problem to be ``naturally'' a QUBO. Note an important consideration for the quantum versus classical comparison of ability is that the tests were done with different equations. Simulated annealing does not require a QUBO. So the simulated annealing was solving an equation with an order of magnitude fewer variables. When simulated annealing was used on the order reduced version, the same version used for quantum annealing, the results were worse and slower than the hybrid method. Ramsey theory beyond $R(3)$ appears
to naturally be of higher order than $2$. This leads to a dramatic increase in variables when converting to quadratic. The simulated annealing required only $105$ variables when searching for $K_4$ monochromatic-free colorings of $K_{15}$. Juxtaposing this, the proposed method when reducing the problem to a QUBO when working with $K_{15}$ required $2967$ variables. The actual required number of variables for this method will be smaller as neither the pre-coloring idea  nor the proposed variable reduction technique was optimized in the implementation, though how much smaller is unclear. The point is that in its implemented form the number of ancillary variables added to reduce degree is an order of magnitude greater than the number of variables that were used to model the problem. As such, without more advanced hardware in terms of qubit count and connectivity, an advancement in the theory, or some novel quantum-classical approach
it is unlikely that QAs could be used to push the bounds of classical Ramsey Theory. 
Using the current methodology to compute a maximal $R(4)$ graph would require quantum hardware with the capacity for a fully connected graph of 1000 logical qubits. This number was determined as a rough proportional estimate scaling from the number of variables required for $K_{13}$. Note also that if one wanted to use quantum annealing to compute $R(5)$ or beyond at this moment using the techniques described in this paper, this would be less efficient asymptotically, with respect to the $K_m$ for which we are avoiding monochromatic $K_n$'s in, than the standard order reduction technique described in the Introduction (section 1). Although future improvements can possibly be made to this method to improve its chances at computing lower bounds for $R(n)$. Additionally, while asymptotically worse, it is unclear how the two methods compare for relevant $m$'s in $K_m$ which in practice are substantially less than infinity.

A more immediate success story lies in the MCT algorithm which was a lynchpin to the Ramsey Theory algorithm. As mentioned earlier, the results produced for minimal triangle colorings of complete graphs matched known solutions. In addition an interesting observation was made when looking at minimal triangle colorings versus known maximal $R(5)$ colorings. The true minimum number of monochromatic triangles were surprisingly close to the number of monochromatic triangles in those maximal $R(5)$ colorings. The minimal number of triangles in a coloring of $K_{42}$ is $2660$ and all the maximal $R(5)$ colorings of $K_{42}$ were within $15$ of this number. Much of the success of the MCT algorithm is attributed to the ability to convert an MCT problem to a QUBO without introducing any auxiliary variables.  


\section{Acknowledgments}

We acknowledge the Institutional Computing (IC) QCAP program at LANL for use of their D-Wave quantum computing allocation. We also
acknowledge the use of the D-Wave Leap Advantage quantum computing resources. Assigned: Los Alamos Unclassified Report
LA-UR-23-32688.

\section{Author Contributions}
J.E.P. and S.M.M. designed the project. J.E.P. performed the numerical simulations and optimizations. S.M.M. supervised the whole project. All authors contributed to the discussion, analysis of the results and the writing of the manuscript.

\section{Funding}
This research was supported by the U.S. Department of Energy
(DOE) National Nuclear Security Administration (NNSA) Advanced
Simulation and Computing (ASC) Beyond Moore's Law (BNL) program at Los Alamos National Laboratory (LANL). This research has also been funded by the LANL Laboratory Directed Research and Development (LDRD) under project number 20230546DR. JEP and SMM were funded by LANL LDRD. SMM was also funded by ASC BML.
Assigned: Los Alamos Unclassified Report LA-UR-23-32688.
LANL is operated by Triad National Security, LLC, for the National
Nuclear Security Administration of U.S. Department of Energy (Contract No. 89233218NCA000001). The funders had no role in study
design, data collection and analysis, decision to publish, or preparation of the manuscript.

\section{Availability of data and materials} All author-produced code will be available upon reasonable request

\bibliography{sn-bibliography}

\end{document}